%% file: ASPDAC24-SoC-Explorer.tex
\documentclass[9pt,conference]{IEEEtran}
\IEEEoverridecommandlockouts 
\input{setting-ieee}

\usepackage{times}
\usepackage{listings}
\graphicspath{{./}}

\begin{document}
\date{}

\title{
    SoC-Tuner: An Importance-guided Exploration Framework for DNN-targeting SoC Design
}

\iftrue
\author{
    Shixin Chen \quad
    Su Zheng \quad
    Chen Bai \quad
    Wenqian Zhao \quad
    Shuo Yin \quad
    Yang Bai \quad
    Bei Yu\\
    The Chinese University of Hong Kong
}
\fi

\maketitle
\pagestyle{plain}

\input{doc/abstract}
\input{doc/intro}

\input{doc/prelim}
\input{doc/soc-tuner}

\input{doc/exp}
\input{doc/conc}

{
    \small
    \bibliographystyle{IEEEtran}
    \bibliography{ref/top-sim,ref/reference}
}

\end{document}

%% file: setting-ieee.tex
\usepackage{blkarray}                                      
\usepackage{algpseudocode}                                 
\usepackage{algorithm}

\usepackage{graphicx}                                      
\usepackage{amsmath}
\usepackage{amssymb}
\usepackage{amsfonts}
\usepackage{amsthm}
\usepackage[mathcal]{eucal}
\usepackage{mathrsfs}
\usepackage{booktabs}
\usepackage{enumerate}
\usepackage{multirow}
\usepackage[subrefformat=parens,farskip=0pt,justification=centering]{subfig}
\usepackage{subfig}
\usepackage{graphicx}
\usepackage{color}
\usepackage{cite}                                          
\usepackage{comment}                                       
\usepackage{soul}                                          
\soulregister\cite7
\soulregister\ref7
\soulregister\pageref7
\usepackage{etoolbox}                                      
\usepackage{url}
\usepackage{nth}                                           
\usepackage{bm}                                            
\usepackage{courier}
\usepackage{balance}
\usepackage{threeparttable}
\usepackage{xcolor,colortbl}
\usepackage{footnote}
\usepackage[bookmarks=false]{hyperref}
\hypersetup{
    colorlinks = true,
    citecolor  = blue,
    linkcolor  = blue,
    urlcolor   = blue,
}
\usepackage{tikz}
\usetikzlibrary{patterns,snakes}
\usetikzlibrary{positioning,calc,fit,decorations.pathmorphing,shapes.geometric, shapes.gates.logic.US, calc}
\usetikzlibrary{arrows,arrows.meta,decorations.markings,shapes,shapes.arrows}
\usetikzlibrary{decorations,decorations.pathreplacing}
\usetikzlibrary{backgrounds}
\usepackage{filecontents}                                  
\usepackage{pgfplots}
\usepackage{pgfplotstable}
\usepackage{scalefnt}
\pgfplotsset{compat=newest}
\usepackage{caption}
\usepackage{cleveref}
\Crefformat{figure}{Fig.~#2#1#3}                           
\Crefname{subfigure}{Fig.}{Figs.}
\Crefname{figure}{Fig.}{Figs.}
\Crefformat{table}{TABLE~#2#1#3}                           
\usepackage[figuresright]{rotating}

\definecolor{CUHKorange}{RGB}{244,106,18} 
\definecolor{CUHKblue}{RGB}{0,111,190}    
\definecolor{CUHKgreen}{RGB}{0,127,128}   
\definecolor{CUHKred}{RGB}{228,46,36}     
\definecolor{CUHKyellow}{RGB}{198,148,34} 
\definecolor{CUHKdark}{RGB}{114,44,114}   
\definecolor{CUHKmiddle}{RGB}{144,44,144} 
\definecolor{CUHKlight}{RGB}{167,44,167} 
\definecolor{CUHKpurple}{RGB}{117,15,109}
\definecolor{CUHKgold}{RGB}{221,163,0}
\definecolor{CUHKribbon}{RGB}{244,223,176}
\definecolor{CUHKblack}{RGB}{34,24,21}

\renewcommand{\bf}[1]{\textbf{#1}}
\renewcommand{\tt}[1]{\texttt{#1}}
\renewcommand{\vec}[1]{\boldsymbol{#1}}    

\newcommand{\minisection}[1]{\vspace{.06in}\noindent{\textbf{#1}}}
\DeclareMathOperator*{\argmin}{argmin}
\DeclareMathOperator*{\argmax}{argmax}

\usepackage{tcolorbox}
\tcbuselibrary{skins,breakable}
    {\endtcolorbox}

\usepackage[top=0.75in,bottom=0.80in,left=0.58in,right=0.58in]{geometry}
\newtheorem{myproblem}{\textbf{Problem}}
\newtheorem{mydefinition}{\textbf{Definition}}

\crefname{mytheorem}{Theorem}{Theorems}
\crefname{mylemma}{Lemma}{Lemmas}
\crefname{myclaim}{Claim}{Claims}
\crefname{myproperty}{Property}{Properties}
\crefname{mycorollary}{Corollary}{Corollaries}

\algrenewcommand\textproc{\texttt}

\makeatletter
\let\OldStatex\Statex
\renewcommand{\Statex}[1][3]{%
  \setlength\@tempdima{\algorithmicindent}%
  \OldStatex\hskip\dimexpr#1\@tempdima\relax
}
\makeatother

\RequirePackage[normalem]{ulem} 
\RequirePackage{color}\definecolor{RED}{rgb}{1,0,0}\definecolor{BLUE}{rgb}{0,0,1} 

%% file: doc/abstract.tex
\begin{abstract}
Designing a system-on-chip (SoC) for deep neural network (DNN) acceleration requires balancing multiple metrics such as latency, power, and area. 
However, most existing methods ignore the interactions among different SoC components and rely on inaccurate and error-prone evaluation tools, leading to inferior SoC design. 
In this paper, we present SoC-Tuner, a DNN-targeting exploration framework to find the Pareto optimal set of SoC configurations efficiently. 
Our framework constructs a thorough SoC design space of all components and divides the exploration into three phases. 
We propose an importance-based analysis to prune the design space, a sampling algorithm to select the most representative initialization points, and an information-guided multi-objective optimization method to balance multiple design metrics of SoC design. 
We validate our framework with the actual very-large-scale-integration (VLSI) flow on various DNN benchmarks and show that it outperforms previous methods.
To the best of our knowledge, this is the first work to construct an exploration framework of SoCs for DNN acceleration.

\end{abstract}

%% file: doc/intro.tex
\section{Introduction}\label{sec:intro}

Designing system-on-chip (SoC) for deep neural network (DNN) acceleration is getting increasingly critical and challenging. 
To keep up with the rapid evolution of the DNN algorithm, the demand for the optimization of DNN accelerators increases as well. 
Although many approaches have been proposed over the past few decades to optimize accelerator design, they may suffer from the rapidly growing scale and complexity and can not perform as effectively or efficiently on the advanced SoC design.  

The first challenge is accurate performance evaluation. 
For example, some existing works targeting accelerators~\cite{ISPASS-2022-VAESA, mei2021zigzag, zhao2023high, bai2021iccad, bai2023tcad} ignored discussions on the interaction between the host processor and the accelerator in SoCs, ignoring the costs of communication and control. 
Therefore, the overall inference latency of DNN given by these tools may be inaccurate, which hinders the design of an optimal accelerator. 
Meanwhile, some analytical tools\cite{samajdar2018scale-sim,parashar2019timeloop, kwon2019maestro} are proposed to evaluate SoC design swiftly. 
However, these tools only consider very limited parameters of SoC architecture and rigidly report the calculation of single-layer, which brings huge gaps to the reporting values and the actual metrics. 
We tackle this problem by implementing the complete very-large-scale-integration (VLSI) flow for authentic and detailed evaluation.

The following challenge is the exploration difficulty, which is rather critical as the SoC design gets more complicated. 
It usually requires many rounds of improvement iterations~\cite{boom-explorer} with domain expertise to get optimal design. 
Time costs will be enormous if we avoid error-prone simple analytical models, as mentioned above. 
Apart from that, The exploration process is heavily dependent on the personal experience of architects, which may bring personal bias in design optimization and result in inferior SoC design.

To bridge the bottlenecks mentioned above, we propose SoC-Tuner, an importance-guided exploration framework to find the optimal SoC design for DNN acceleration. 
SoC-Tuner aims to find the optimal SoC design, which balances multiple metrics including inference latency, power consumption, and chip area for various DNNs. 
Our contributions can be concluded as follows:
\begin{itemize}
    \item We thoroughly consider various SoC components that influence DNN computations and construct a huge design space to avoid insufficient evaluation of overall DNN inference. 
    \item We employ actual very-large-scale-integration (VLSI) flow to evaluate multiple metrics, which achieves more accurate modeling of SoC than simplified analytical tools. 
    \item We propose an importance-based analysis to prune the design space, a sampling algorithm to select the most representative initialization points, and an information-guided multi-objective optimization method to balance multiple design metrics of SoC design.

    \item Experimental results demonstrated the efficiency and effectiveness of our framework on various benchmarks compared to some state-of-the-art methods.
\end{itemize}


%% file: doc/prelim.tex
\section{Preliminaries}\label{sec:Prelim}

\subsection{SoC with DNN Accelerator}

\begin{figure}[t]
\centering
	 \centering
    \includegraphics[width=0.96\linewidth]{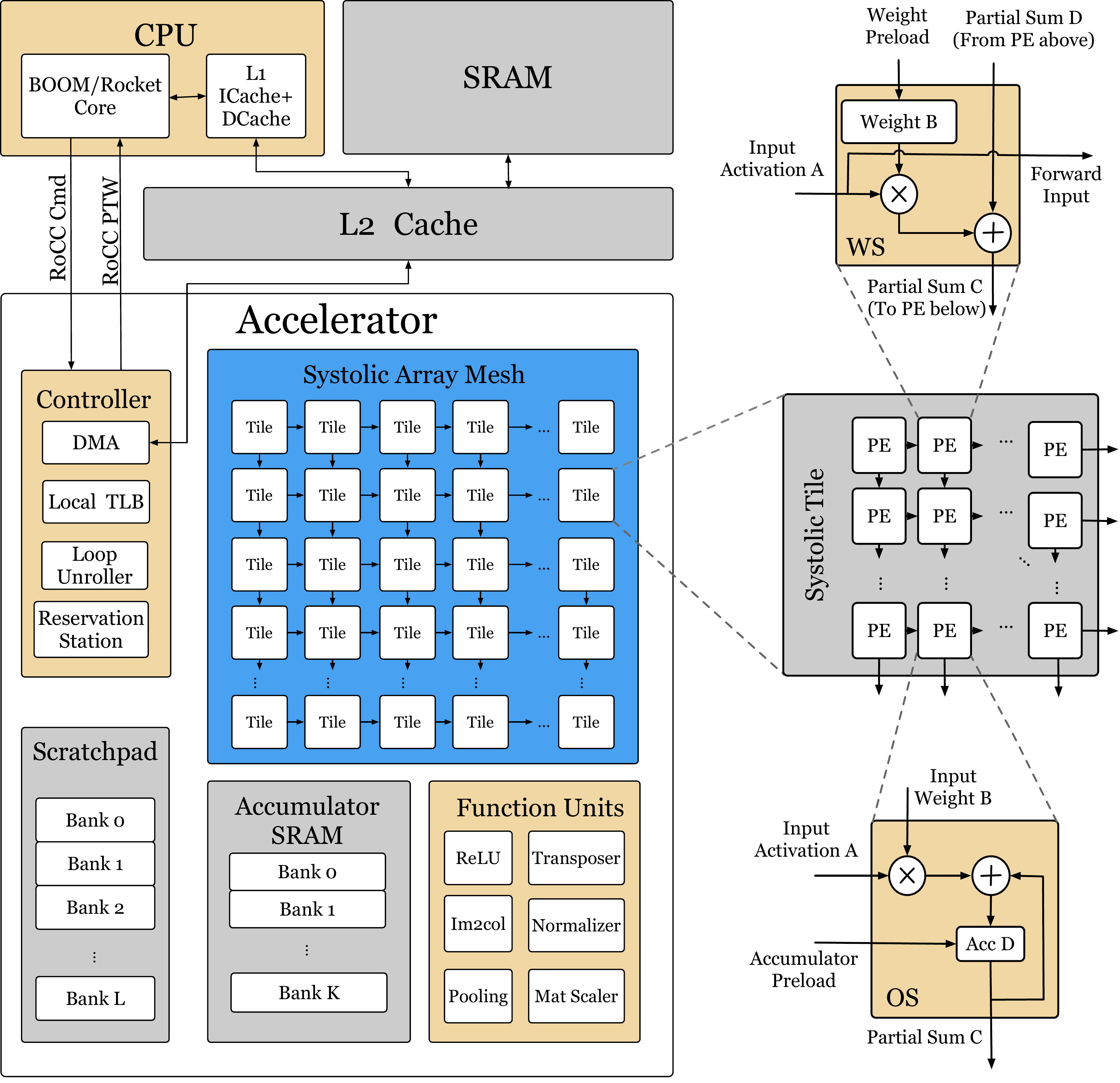}
    \caption{Architecture of a SoC with a systolic-based accelerator.  
    }
    \label{fig:gemmini architecture}
\end{figure}

A typical DNN-targeting SoC containing an accelerator is shown in \Cref{fig:gemmini architecture}, where the systolic array\cite{chen2016eyeriss}\cite{jouppi2017tpu} is one of the most widely used architectures for DNN accelerators.
In \Cref{fig:gemmini architecture}, the CPU allocates instructions to the accelerator using Rocket\cite{asanovic2016rocket} co-processor command (RoCC) instructions including \texttt{Load}, \texttt{Store}, and \texttt{Execute}.
The SRAM stores the DNN models to be computed by the accelerator. 
The CPU and accelerator share L2 cache. 

To facilitate designing SoC rapidly, we employ the agile hardware development method.
It aims to construct highly modeled and parameterized hardware components in Chisel\cite{bachrach2012chisel} language, which can be easily initiated with various architecture parameters. 
Chipyard\cite{amid2020chipyard} is integration with a variety of hardware components like  CPU cores (\textit{e.g.}, in-order Rocket\cite{asanovic2016rocket} and out-of-order CPU core BOOM\cite{asanovic2015BOOM}) and co-processors (\textit{e.g.}, vector-thread processor Hwacha\cite{lee2015hwacha} and 
DNN accelerator Gemmini\cite{genc2021gemmini}). 
It provides us an opportunity to easily design new SoC architectures given existing components.

Designing an optimal SoC design with given components is time-consuming and complicated.
Previous work\cite{boom-explorer} explores the design space of microarchitecture of processor core with learning methods, and \cite{chen2014archranker} utilizes a ranking-based approach to explore the optimal design of CPU core.
However, these works solely focus on a single processor and bring challenges to more complicated systems like SoC. 
In SoC design, the control logic and computations are rather complicated due to the interactions between various components shown in \Cref{fig:gemmini architecture}. 
Moreover, the design space of SoC is more huge than a single CPU.
Therefore, an exploration framework for characteristics of SoC design is necessary.

\subsection{Problem Formulation}

\begin{mydefinition}[SoC Architecture Design]
A combination of the features listed in TABLE \ref{tab:gemmini-design-space} is denoted as a design point $\vec{x}$ of SoC, and all design points make up the entire design space $\mathcal{X}$. 
An SoC architecture can be determined by a design point $\vec{x}$. 
We define the SoC design problem as finding an $\vec{x} \in \mathcal{X}$ to design an SoC that balances the latency, power, and area for various DNNs.
\end{mydefinition}

\begin{mydefinition}[Multi-objectives Optimization of SoC]
    Multi-objective optimization of SoC is defined to find $\vec{x} \in \mathcal{X}$ to trade-off $m$ objective functions $\{f_{1}(\vec{x}), \cdots , f_{m}(\vec{x})\}$. 
    We define $\vec{y}=\mathbb{F}(\vec{x})=(y_{1},\cdots, y_{m})$, where different $y_{i}=f_{i}(\vec{x})$ is obtained by various evaluation tools based on $\vec{x}$.
    All $\Vec{y}$ formulates metrics space $\mathcal{Y}=\{\vec{y}|\vec{y}=\mathbb{F}(\vec{x}) ,\vec{x} \in \mathcal{X}\}$.
\end{mydefinition}

\begin{mydefinition}[Pareto Optimal Set of SoC]
    For an optimization problem, an $m$-dimensional objective $\vec{y}=\mathbb{F}(\vec{x})$ is said to be dominated by $\vec{y}^{\ast}=\mathbb{F}(\vec{x}^{\ast})$ if
    \vspace{-0.0cm}
    \begin{equation} \begin{aligned}
        &\forall i \in [1, m], \mathbb{F}_{i}(\vec{x})\le \mathbb{F}_{i}(\vec{x}^{\ast}); \\
        &\exists j \in [1, m], \mathbb{F}_{j}(\vec{x}) < \mathbb{F}_{j}(\vec{x}^{\ast}),
    \end{aligned} \end{equation}
    where we denote $\vec{y}^{\ast} \succcurlyeq \vec{y}$ to represent this situation.
    In the entire design space, a set of design points not dominated by any other points form the Pareto optimal set.
    In the Pareto optimal set, a design point can not be optimized without sacrificing other objectives.
\end{mydefinition}

In the design space exploration of SoC, the chip area, inference latency, and power consumption are a group of negatively correlated metrics, so an SoC design cannot improve one metric without sacrificing another metric. 
Therefore, to design an optimal SoC is to find the Pareto optimal set of SoC designs, and then choose one design point that balances multiple objectives.

\begin{myproblem}[Design Space Exploration of SoC]
    Given an SoC design space $\mathcal{X}$, 
and the metrics space $\mathcal{Y}=\{\vec{y}|\vec{y}=\mathbb{F}(\vec{x}) ,\vec{x} \in \mathcal{X}\}$,
we define the design space exploration of SoC as finding a subset $\mathcal{X}^{\ast} \in \mathcal{X}$, whose corresponding metrics $\mathcal{Y}^{\ast}$ form the Pareto optimal set. 
Hence,
\vspace{-0.0cm}
\begin{equation}
    \begin{aligned}  
        \mathcal{Y}^{\ast}& = \{\vec{y}^{\ast}| \vec{y}^{\ast}\nsucceq \vec{y}, \forall \vec{y} \in \mathcal{Y} \},\\ \mathcal{X}^{\ast} &= \{\vec{x}| \mathbb{F}(\vec{x}) \in \mathcal{Y}^{\ast}, \forall \vec{x} \in \mathcal{X}\}.
    \end{aligned}
\end{equation}
\end{myproblem}

%% file: doc/soc-tuner.tex
\section{Methodology}\label{sec:SoC}
\subsection{Overview of SoC-Tuner}

\begin{figure}[t!]
    \centering
    \includegraphics[width = 0.96\linewidth]{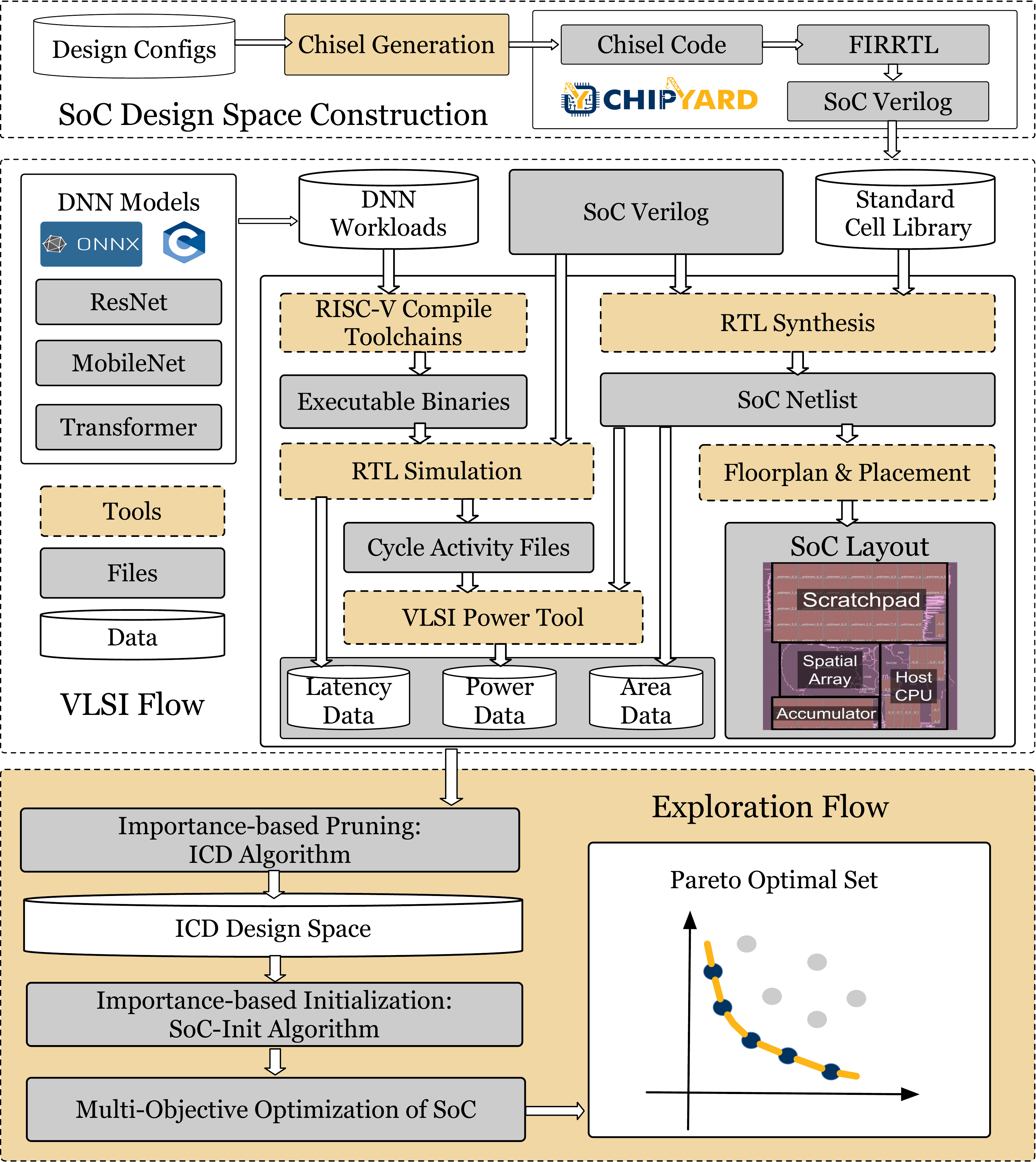}
    \caption{The overall flow of the proposed SoC-Tuner.}
    \label{fig:soc-framework}
\end{figure}

The overall flow of our framework is displayed in \Cref{fig:soc-framework}, which contains three parts, including SoC Design Space Construction, VLSI Flow, and Exploration Flow.

In SoC Design Space Construction, we implement a Chisel generation tool to take in design points from design configurations and generate Verilog-based SoC design in Chipyard. 
Then we utilize the VLSI Flow that consists of several tools colored in yellow and intermediate files colored in gray. 
Its outputs are important metrics like latency, power, and chip area of SoC. 
To evaluate the performance on DNN workloads, popular DNNs like ResNet, MobileNet, Transformer, \textit{etc.} are provided in open neural network exchange (ONNX) format or C code.
In the Exploration Flow, the metrics data from the VLSI Flow are fed in to optimize the SoC design. 
We propose an inter-cluster distance (ICD) algorithm for design space pruning, and an SoC-Init algorithm for exploration initialization, which improve the efficiency of exploration efficiency of huge design space. 
Finally, we employ correlated multi-objective Bayesian optimization to find the Pareto optimal set of SoC designs that balance multiple metrics.
The details of SoC-Tuner will be elaborated as follows.
\begin{table}[t]
    \centering
    \caption{Selected parameters from the SoC containing a systolic-based accelerator }
    \label{tab:gemmini-design-space}
    \resizebox{0.93\linewidth}{!}
    {
        \begin{tabular}{c|c|c}
          \toprule
           {Components} & {Descriptions} & {Candidate Values} \\
          \midrule
          \midrule
          HostCore          & Various Host CPU core                                   & c1, c2, c3\\
          L2Bank            & Entries of L2 cache banks                               & 1, 2, 4\\
          L2Way             & Entries of L2 cache way                                 & 4, 8, 16 \\
          L2Capa            & Capacity of L2 cache bank                  & 128, 256, 512\\
          \midrule
          Tilerow/col       & Dimension of the tile                  & 1, 2, 4, 8\\   
          Meshrow/col       & Dimension of the mesh        & 8, 16, 32, 64 \\
          Dataflow          & Dataflow mode of systolic array                 & WS, OS, BOTH \\
          InputType         & Bit width of input datatype                     & 8, 16, 32\\
          AccType           & Bit width of accumulator datatype                & 8, 16, 32     \\
          OutType           & Bit width of output datatype                    & 8, 20, 32      \\
          \midrule
          SpBank            & Banks of scratchpad memory                      & 4, 8, 16,32 \\
          SpCapa            & Entries of scratchpad bank                  & 64, 128, 256, 512 \\
          AccBank           & Banks of accumulator memory                     & 1, 2 ,4 ,8 \\
          AccCapa           & Entries of accumulator bank                 & 64, 128, 256, 512 \\
          \midrule
          LdQueue           & Entries of the \texttt{Lord} queue         & 2, 4, 8, 16 \\
          StQueue           & Entries of the \texttt{Store} queue        & 2, 4, 8, 16 \\
          ExQueue           & Entries of the \texttt{Execute} queue      & 2, 4, 8, 16 \\
          LdRes             & Entries of the \texttt{Lord} in ROB        & 2, 4, 8, 16 \\
          StRes             & Entries of the \texttt{Store} in ROB       & 2, 4, 8, 16 \\
          ExRes             & Entries of the \texttt{Execute} in ROB     & 2, 4, 8, 16 \\
          \midrule
          MemReq            & memory requests in-flight             & 16, 32, 64\\
          DMABus            & Width of DMA bus                                & 32, 64, 128\\
          DMABytes          & Number of bytes in DMA bus                      & 32, 64, 128 \\
          TLBSize           & Size of TLB page                                & 4, 8, 16     \\
          \bottomrule
        \end{tabular}
    }
\end{table}

\subsection{SoC Design Space Construction}

In \Cref{fig:gemmini architecture}, the blue part shows the detailed structure of a mesh of tiles in the systolic array. 
A tile is an array consisting of a grid of processing elements (PEs) that can perform parallel multiplication-accumulation (\texttt{MAC}): $C=A \times B + D$, where $A, B, D$ represent the activation matrix, the weight matrix, and the result of the prior \texttt{MAC}, respectively. 
The right part of \Cref{fig:gemmini architecture} illustrates the details of these two modes of the systolic array, \textit{i.e.}, weight-stationary (WS), and output-stationary (OS). 
In WS mode, the weight of DNNs is pre-stored in the PEs, while in OS mode, the partial sum of computations is pre-stored in the systolic array.  
We can choose either mode or both of them in an SoC design depending on different DNNs.

By thoroughly considering all components that influence the metrics of SoC, we build the \Cref{tab:gemmini-design-space} that lists the design parameters of the whole SoC in \Cref{fig:gemmini architecture}.
All combinations of features in \Cref{tab:gemmini-design-space} form a huge design space of all possible SoC designs. 
The SoC parameters are classified into several groups \textit{i.e.}, CPU core \& L2 cache, systolic array, accelerator memory, accelerator controller, and RoCC communication shown from the top to the bottom of \Cref{tab:gemmini-design-space}. 
Three representative CPU cores, \textit{i.e.}, c1 (\texttt{LargeBoom}), c2 (\texttt{LargeRocket}), and c3 (\texttt{MedRocket}) are chosen as the candidates of the host core.
For the 

\subsection{Importance-based Pruning and Initialization}

\minisection{Importance-based SoC Design Space Pruning}.
Designing an SoC with high performance is complicated and time-consuming. 
Developers should choose the most representative design points to evaluate the SoC design and get adequate information to guide the design. 
Due to the time-consuming VLSI flow, only a limited number of designs will be synthesized to obtain evaluation metrics. 
However, randomly sampling the design parameters like \cite{lee2007regressDSE} may ignore some domain-specific knowledge in SoC design. 

In fact, there exist important features that have a significant influence on the metrics of the SoC, which means that by slightly changing the feature value, the metrics will change heavily. 
To model this influence, we use a vector $\vec{v}$ to denote the importance of each parameter and propose \Cref{alg:ICD} to evaluate the parameter importance via a few VLSI flow trials. 
In \Cref{alg:ICD}, line 1 represents a few VLSI trials, and line 4 clusters the metrics space $\mathcal{Y}^{\prime}$ into several groups according to candidates of various design features. 
Line 5 $\sim$ 8 is to get the average metric vectors of each group. 
Line 9 illustrates the conclusion of inter-cluster distance (ICD), where $C_{2}^{|M|}$ is the number of two-combination in average vectors $M$. Finally, after normalization, a $d_{x}$-dimension vector $\vec{v}$ is given to represent the importance of each design feature.

\begin{figure}[tb!]
    \center
    \includegraphics[width = 0.90\linewidth]{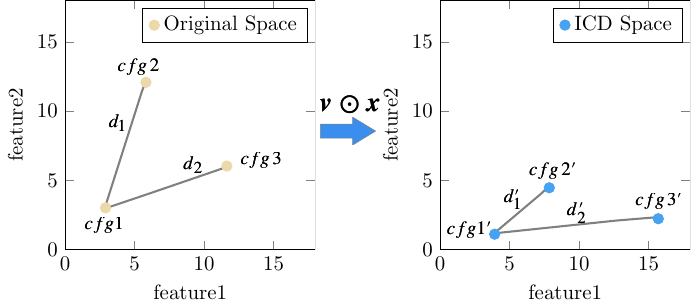}
    \caption{A toy example with two features shows the transformation from the original space to the ICD space.}
    \label{fig:ICD-space}
\end{figure}
\begin{algorithm}[tb!]
    \small
	\renewcommand{\algorithmicrequire}{\textbf{Input:}}
	\renewcommand{\algorithmicensure}{\textbf{Output:}}
    \caption{\texttt{ICD} ($\mathcal{X},n$)}
	\label{alg1}
	\begin{algorithmic}[1]
	    \Require ($\mathcal{X},n$), where $\mathcal{X}$ is the whole design space, $n$ is the trial times of importance analysis. In a $d_{x}$-dimension design point $\vec{x}=\{x_1,x_2,...,x_{d_x}\} \in \mathcal{X}$, each feature $x_{i}, i \in \{1, \cdots,d_{x}\}$ has $t_{i}$ design candidates. And $d_y$ is the dimension of the metrics $\vec{y} \in \mathcal{Y}$.
	    
	    \Ensure The feature importance value factor $\vec{v}$.
        \State $\mathcal{Y}^{\prime}=\mathbf{VLSIFlow}(\texttt{Sample}(\mathcal{X},n))$;
		\State $\vec{v}=\emptyset$, $M=\emptyset$;
		\For {$i\in \{1,2,\cdots,{d_x}\}$} 
                \State $\{\mathcal{Y}^{\prime}_1,\cdots,\mathcal{Y}^{\prime}_{t_{i}}\} \leftarrow \mathcal{Y}^{\prime}$;\Comment{clusters based on $t_{i}$ candidates of $x_{i}$.}
		    \For {$j \in  \{1,2,\cdots,t_{i}\}$}
                \State $m_{j}=\texttt{mean}(\mathcal{Y}^{\prime}_{j})$; \Comment{average vector of $\mathcal{Y}^{\prime}_{j}$.}
                \State $M = M \bigcup \{m_{j}\}$;
		    \EndFor
            \vspace{.1in}
		    \State $v_{i}=\dfrac{\sum_{p,q} ||m_p-m_q||_2}{C_{2}^{|M|}}, p,q \in \{1,\cdots,t_{i}\} $;
		    \State $\vec{v}=\vec{v} \bigcup \{v_i\}$;
		    \EndFor 
        \State $\vec{v}=\texttt{normalize}(\vec{v})$;\\
		    \Return $\vec{v}$;
	\end{algorithmic}
 \label{alg:ICD}
\end{algorithm}

Considering all the features listed in \Cref{tab:gemmini-design-space}, the whole SoC design space $\mathcal{X}$ is too huge to fully explore. 
To avoid unnecessary exploration brought by less important feature parameters, the original design space $\mathcal{X}$ will be pruned based on $\vec{v}$ given by the ICD algorithm. 
Supposing $\{\vec{x}_{i}^{1},\cdots, \vec{x}_{i}^{j} \}$ indicates the $j$ candidates of the $i^{th}$ feature of design point $\vec{x}$, we can use ICD vector $\vec{v}$ to prune the design space, shown in line 1 of \Cref{alg:SoC-Init}, where $v_{th}$ represents the importance threshold. 
Higher $v_{th}$ will remove more design points, and $\text{medium}(.)$ chooses the medium value. 

\minisection{Importance-based SoC Exploration Initialization}.
To take the importance value into initialization, line 2 in \Cref{alg:SoC-Init} uses element-wise multiplication $\odot$ to transform the original design space to ICD space.
In this way, design points with similar influences on metrics will move closer, and points with significant differences in metrics will move more separately. 

\Cref{fig:ICD-space} shows a toy example with 2 design features for transforming from the original space to the ICD space, where feature1 is important and feature2 is less important. 
After importance-based transformation with $\vec{v}$, $cfg2^{\prime}$ will move closer to $cfg1^{\prime}$, while $cfg3^{\prime}$ will move further from $cfg1^{\prime}$. 
In this way, the importance of parameters is introduced to ICD space when uniformly samples design points for initialization of space exploration.
After the exploration, the Pareto optimal set will be transformed into the original space to use the feature parameters to design the optimal SoC.

To summarize the methods mentioned above, we proposed an importance-guided SoC-Init algorithm
shown in \Cref{alg:SoC-Init}. 
The most significant inputs of the algorithm are the original design space, and the ICD values obtained from \Cref{alg:ICD}.
Given the ICD design space $\mathcal{X}^{\prime}$ after pruning (line 1) and space transformation (line 2), the SoC-Init algorithm will sample a subset $\mathcal{Z} \in \mathcal{X}^{\prime}$ for initialization of exploration. 
In line 3, $\vec{K}=\vec{K}_{\mathcal{X}^{\prime}\mathcal{X}^{\prime}} \in \mathbb{R}^{|\mathcal{X}^{\prime}|\times|\mathcal{X}^{\prime}|}$ is the distance matrix of all design points in $\mathcal{X}^{\prime}$, and $\Phi(\vec{x}_{i}^{\prime}, \vec{x}_{j}^{\prime}) \in \vec{K}_{\mathcal{X}^{\prime}\mathcal{X}^{\prime}}, i,j \in \{1,2,\cdots,|\mathcal{X}^{\prime}|\}$ is computed as Euclidean distance, with $\vec{x}_{i}^{\prime}, \vec{x}_{j}^{\prime} \in \mathcal{X}^{\prime}$. 
To make the initial configurations have higher diversity and scatter the whole design space, we use the TED\cite{yu2006TED} method to sample design points from the ICD space. 
The design points that contribute most to initialization will be sampled by \Cref{alg:SoC-Init}.       

\begin{algorithm}[!t]
    \small
	\renewcommand{\algorithmicrequire}{\textbf{Input:}}
	\renewcommand{\algorithmicensure}{\textbf{Output:}}
	\caption{\texttt{SoC-Init}($\mathcal{X},u,b,\vec{v},v_{th}$)}
	\label{alg1}
	\begin{algorithmic}[1]
	    \Require ($\mathcal{X},u,b,\vec{v},v_{th}$), where $\mathcal{X}$ is the un-sampled design space, $\mu$ is the normalization coefficient, $b$ is the number of configurations we will sample, $\vec{v}$ is the ICD feature vector from Algorithm \ref{alg:ICD}. 
	    \Ensure  $\mathcal{Z}$, the sampled set with $|\mathcal{Z}|=b$.
            \State  $\mathcal{Z} = \emptyset$; $\text{if } v_{i} < v_{th},\text{then}\forall \vec{x} \in \mathcal{X}, \vec{x}_{i}=\text{medium}(\{\vec{x}_{i}^{1},\cdots, \vec{x}_{i}^{j} \})$;
            \State  $\mathcal{X}^{\prime}=\{\vec{v} \odot \vec{x}, \forall \vec{x} \in \mathcal{X}\} $; 
            \State $\vec{K}= \{ \Phi (\vec{x}^{\prime}_{i},\vec{x}^{\prime}_{j})| \vec{x}^{\prime}_{i}, \vec{x}^{\prime}_{j} \in \mathcal{X}^{\prime}\}$; \Comment{ $\Phi(.)$ is Euclidean distance.} 
		\For {$i \in \{1,2,\cdots,b\}$} 
		      \State $\vec{z} = \argmax_{\vec{x}^{\prime} \in \mathcal{X}^{\prime}} =\dfrac{|| \vec{K}_{\vec{x}^{\prime}}||^{2}}{\Phi(\vec{x}^{\prime},\vec{x}^{\prime})+\mu}$;
                \Comment{$\vec{K}_{\vec{x}^{\prime}}$ and $\Phi(\vec{x}^{\prime},\vec{x}^{\prime})$ are $\vec{x}^{\prime}$'s corresponding column and diagonal entry in $\vec{K}$.} 
                \State $\mathcal{Z}=\mathcal{Z}\bigcup \{\vec{z}\}$;
                \State $\vec{K}= \vec{K} - \dfrac{\vec{K}_{z} \vec{K}^{\top}_{z}}{\Phi(\vec{z},\vec{z})+\mu}$;
		\EndFor \\
	\Return $\mathcal{Z}$;
	\end{algorithmic}
    \label{alg:SoC-Init}
\end{algorithm}
\subsection{ Multi-Objective Exploration with Information Gain}

Even though we have carefully chosen the initial set $\mathcal{Z}$ via the SoC-Init algorithm, building a model that can mimic the relationship between the configurations and the objectives is not easy. 
Witnessing that the Gaussian process (GP) shows robustness and non-parametric approximation in various domains\cite{sun2021dseAL,ma2018dseadder,zhao2023high},  we choose GP as our surrogate model.

We have the ICD design space $\mathcal{X}^{\prime}$ consisting of design parameters, and according to different $\vec{x}^{\prime} \in \mathcal{X}^{\prime}$, we can get metrics space $\mathcal{Y}$ with time-consuming VLSI Flow shown in \Cref{fig:soc-framework}. 
GP provides a prior over the function $f(\vec{x}^{\prime}) \sim \mathcal{GP}(\bm{\mu}, k_{\bm{\theta}})$, where $\bm{\mu}$ is the mean value and the kernel function $k$ is parameterized by $\bm{\theta}$. All the objective functions (design metrics) can be expressed as a group of GP models and combined as Equation \eqref{equ: GP}.

\vspace{-0.5cm}
\begin{equation}
    \mathbb{F}=[f(\vec{x}_{1}^{\prime}), f(\vec{x}_{2}^{\prime}),\cdots, f(\vec{x}_{n}^{\prime}))]^{T} \sim \mathcal{N}(\bm{\mu}, \vec{K}_{\mathcal{X}^{\prime}\mathcal{X}^{\prime}|\bm{\theta}}),
    \label{equ: GP}
\end{equation}
where $\vec{K}_{\mathcal{X}^{\prime}\mathcal{X}^{\prime}|\bm{\theta}}$ is the intra-covariance matrix among all feature vectors and can be computed via $[\vec{K}_{\mathcal{X}^{\prime}\mathcal{X}^{\prime}|\bm{\theta}}]_{ij}=k_{\bm{\theta}(\vec{x}_{i}^{\prime},\vec{x}_{j}^{\prime})}$, and Gaussian noise $\mathcal{N}(f(\vec{x}^{\prime}),\sigma^{2}_{e})$ is to model uncertainties of GP models. 
Given a newly sampled feature vector $\vec{x}_{\ast}^{\prime}$, the predictive joint distribution $f_{\ast}$ based on $\vec{y}$ is calculated by Equation \eqref{equ:new_sample}.
\begin{equation}
    f_{\ast}|\vec{y} \sim \mathcal{N}( \left [\begin{array}{cc}
         \bm{\mu}\\
         \mu_{\ast} 
    \end{array} \right ], \left[\begin{array}{cc}
\vec{K}_{\mathcal{X}^{\prime}\mathcal{X}^{\prime}|\bm{\theta}}+\sigma^{2}_{e}\mathbf{I}   & \vec{K}_{\mathcal{X}^{\prime}\vec{x}_{\ast}^{\prime}|\bm{\theta}}  \\
       \vec{K}_{\vec{x}_{\ast}^{\prime}\mathcal{X}^{\prime}|\bm{\theta}}  & k_{\vec{x}_{\ast}^{\prime}\vec{x}_{\ast}^{\prime}|\bm{\theta}} 
    \end{array} \right]).
    \label{equ:new_sample}
\end{equation}

By maximizing the marginal likelihood of GP, $\bm{\theta}$ is optimized to sense the entire design space. 
Each time we get $\vec{y}$ from VLSI Flow, $\bm{\theta}$ will be updated to better mimic the complex relationship between the design space $\mathcal{X}^{\prime}$ and metrics space $\mathcal{Y}$. 

Therefore, deciding the next $\vec{x}^{\prime}$ to be sent to the VLSI flow is important to optimize the surrogate model. 
From each $\vec{y}$ obtained from VLSI, we need to maximize the information gained about the Pareto optimal set $\mathcal{Y}^{\ast}$ as much as possible.
So we develop an information gain-based acquisition function $I(\vec{x}^{\prime})$ expressed with entropy $H(\cdot)$ as follows.
\begin{align}
     I(\vec{x}^{\prime})
                &=H(\mathcal{Y}^{\ast} | \mathcal{X}^{\prime})-\mathbb{E}_{\vec{y}}[H(\mathcal{Y}^{\ast} | \mathcal{X}^{\prime} \cup \{\vec{x}^{\prime},\vec{y}\} )]\\
                &=H(\vec{y} | \mathcal{X}^{\prime},\vec{x}^{\prime})-\mathbb{E}_{\mathcal{Y}^{\ast}}[H(\vec{y} | \mathcal{X}^{\prime},\vec{x}^{\prime},\mathcal{Y}^{\ast} ) ]\\
                &\simeq H(\vec{y} | \mathcal{X}^{\prime},\vec{x}^{\prime})-\dfrac{1}{S}\sum^{S}_{s=1}[H(\vec{y} | \mathcal{X}, \vec{x}^{\prime},\mathcal{Y}^{\ast}_{s})],
                \label{equ:acq}
\end{align}
where \Cref{equ:acq} is approximately computed via Monte-Carlo sampling, and $S$ is the number of samples and $\mathcal{Y}^{\ast}_{s}$ denote a sampled Pareto optimal set.

\begin{algorithm}[t]
    \small
	\renewcommand{\algorithmicrequire}{\textbf{Input:}}
	\renewcommand{\algorithmicensure}{\textbf{Output:}}
	\caption{\texttt{SoC-Tuner}($\mathcal{X},T,n,u,b,v_{th}$)}
	\label{alg1}
	\begin{algorithmic}[1]
	    \Require $\mathcal{X}$ is the unsampled SoC design space, $T$ is the maximal iteration number of BO, $n$ is the trail times of importance analysis, $\mu$ is the normalization coefficient, $b$ is the number of samples for initialization.
	    \Ensure  The Pareto optimal set $\mathcal{X}^{\ast}$ and corresponding $\mathcal{Y}^{\ast}$.
            \State  $\vec{v}=\texttt{ICD}(\mathcal{X},n)$;\Comment{Algorithm \ref{alg:ICD}.}
	    \State  $\mathcal{Z}=$\texttt{SoC-Init}$(\mathcal{X},\mu,b,\vec{v},
        v_{th})$;\Comment{Algorithm \ref{alg:SoC-Init}.}
            \State  $\mathcal{X}^{\prime}=\{\vec{v} \odot \vec{x},\forall \vec{x} \in \mathcal{X}\} $; 
	    \State $\vec{y}$ $\leftarrow$ $\texttt{VLSIFlow}(\mathcal{Z})$;
	    \For {$i \in \{1,2,...,T\}$}
                \State Construct the Pareto optimal set $\mathcal{Y}^{\ast}$ from $\vec{y}$;
	        \State $\vec{x}^{\ast}$ $\leftarrow$ \texttt{IMOO}$(\mathcal{X}^{\prime},\mathcal{Y}^{\ast},\bm{\theta})$;\Comment{Equation \eqref{equ:pac-moo}.}
	        \State $\mathcal{Z}=\mathcal{Z} \bigcup \{\vec{x}^{\ast}\}$, $\vec{y} = $ $\vec{y}\bigcup \{\texttt{VLSIFlow}(\vec{x}^{\ast})\}$;
                \State $\bm{\theta}$ is optimized via gradient descent method;
	    \EndFor
	    \State Construct Pareto optimal set $\mathcal{Y}^{\ast}$ from $\mathcal{Z}$, and restore the corresponding $\mathcal{X}^{\ast}$ from the ICD space;\\
        \Return $\mathcal{X}^{\ast}$
	\end{algorithmic}
 \label{alg: over all alg}
\end{algorithm}

 The value of each element of $\vec{y}$ in \Cref{equ:acq} is upper bounded by the maximum value of the corresponding element in sampled point on Pareto optimal set $\mathcal{Y}^{\ast}$. We can combine the boundedness property and the fact that each sampled objective function is modeled as a GP prior, and treat each component of $\vec{y}$ as a truncated Gaussian distribution. Then we can rewrite Equation \eqref{equ:acq}, and obtain the approximation of the acquisition function,
 \begin{align}
     AF(i,\vec{x}^{\prime})      &=\sum^{S}_{s=1}\dfrac{\gamma^{i}_{s}(\vec{x}^{\prime})\phi(\gamma^{i}_{s}(\vec{x}^{\prime}))}{2\phi(\gamma^{i}_{s}(\vec{x}^{\prime}))}-ln(\phi(\gamma^{i}_{s}(\vec{x}^{\prime}))),\\
    I(\vec{x}^{\prime}) &\simeq \sum_{i \in \mathcal{I}} AF(i,\vec{x}^{\prime}), \mathcal{I}=\{f_{1},\cdots, f_{n}\}, 
 \end{align}
 where $\gamma$ and $\phi$ stand for the probability density function and the cumulative density function of a standard Gaussian distribution, respectively. $\gamma^{i}_{s}(\vec{x}^{\prime})$ equals $\dfrac{\vec{y}^{\ast}_{s}-\mu_{s}(\vec{x}^{\prime})}{\sigma_{s}(\vec{x}^{\prime})}$ with $\vec{y}^{\ast}_{s}$ is the maximum value among the sampled points on predicted Pareto optimal set for the $i^{th}$ objective.

 Ultimately, we can choose $\vec{x}^{\ast}$ that maximizes the Equation \eqref{equ:final-acq} as the next design point sent to VLSI flow:
 \begin{align}
   \vec{x}^{\ast} = \argmin_{\vec{x}^{\prime}} I(\vec{x}^{\prime}),
   \label{equ:final-acq}
 \end{align}
 where $\vec{x}^{\ast}$ will bring the most information gain.
 To conclude Equation \eqref{equ: GP} to \eqref{equ:final-acq}, we can design an information-gain-based multi-objective optimization (IMOO):
 \begin{align}
   \vec{x}^{\ast} \leftarrow \texttt{IMOO}(\mathcal{X}^{\prime},\mathcal{Y}^{\ast},\bm{\theta}),
   \label{equ:pac-moo}
 \end{align}
where $\mathcal{X}^{\prime}$ is ICD Space given by \Cref{alg:SoC-Init}, $\mathcal{Y}^{\ast}$ is the current Pareto optimal set,
and $\bm{\theta}$ is optimized to better mimic the surrogate GPs model. 
We combine all the proposed algorithms into the overall algorithm illustrated in \Cref{alg: over all alg}.
The maximal number of exploration rounds is $T$, and the output of SoC-Tuner is the learned Pareto optimal set.

%% file: doc/exp.tex
\begin{figure*}[tb!]
    \centering
    \subfloat[Learned Pareto optimal set \\(inference latency \textit{vs} chip area)]{
        \includegraphics[width=.34\textwidth]{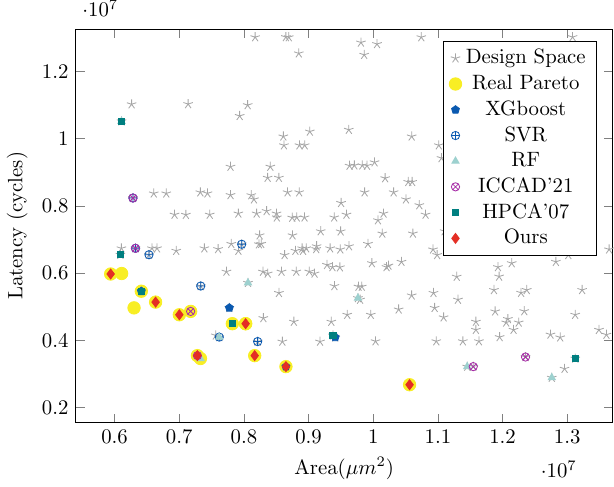}  \label{fig:pareto-a}
    }
    \subfloat[Learned Pareto optimal set \\(inference latency \textit{vs} power consumption) ]{
        \includegraphics[width=.35\textwidth]{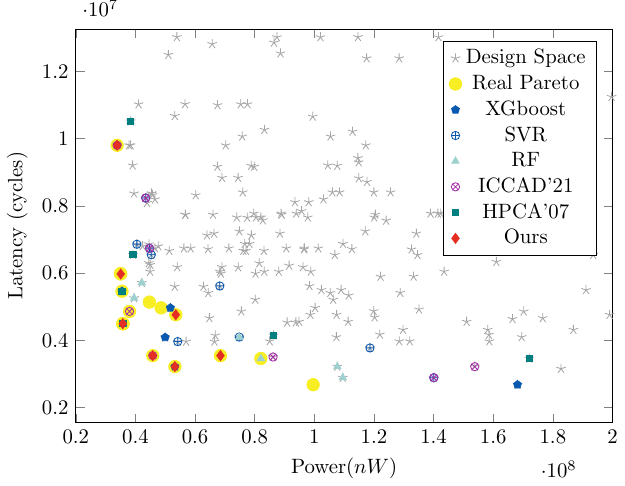} \label{fig:pareto-b}
    }
    \subfloat[Simplified model ($^\ast$actual metrics, $^\dag$ reported metrics) \textit{vs} RTL-simulation ]{
        \includegraphics[width=.26\textwidth]{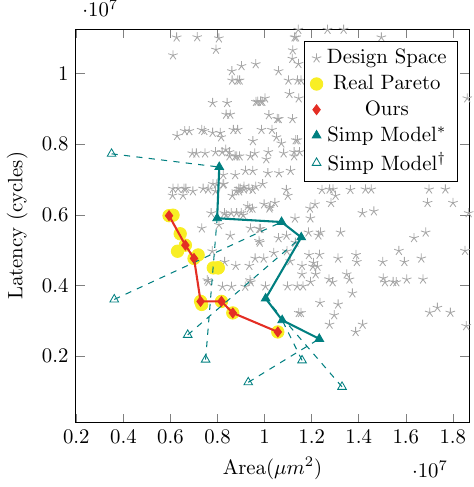} \label{fig:pareto-c}
    }
    \caption{
        The Pareto optimal set is given by various methods (ResNet50).
        (a) and (b) demonstrate the effectiveness of our framework.
        (c) shows that the metrics from the simplified model have a big gap with that from the VLSI flow.}
    \label{fig:pareto} 
\end{figure*}

\begin{figure}[tb!]
    \centering
    \includegraphics[width = 0.48 \textwidth]{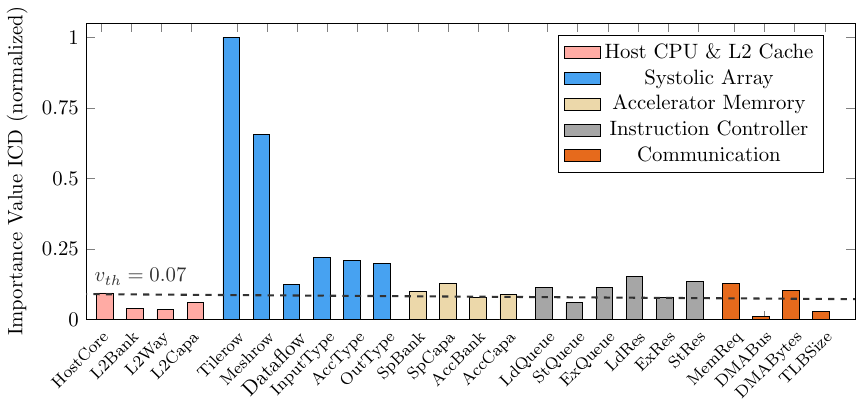}
    \caption{Importance analysis of the parameters given by ICD algorithm($n=30,v_{th}=0.07$).}
    \label{fig:importance}
\end{figure}

\section{Experiment \& Analysis}\label{sec:exp}



\subsection{Benchmarks and Baselines}
To evaluate the SoC design uniformly, we randomly sample 2500 design points in \Cref{tab:gemmini-design-space}. 
Each design is evaluated with the VLSI flow in \Cref{fig:soc-framework}, and the metric data are collected to verify various methods. 
We choose some popular DNNs including ResNet50\cite{he2016resnet}, MobileNet\cite{howard2017mobilenets}, and Transformer\cite{vaswani2017transformer} as our benchmarks. 

Several representative baselines are compared with SoC-Tuner. 
The MicroAL-based method\cite{boom-explorer} (ICCAD'21) is used to predict the power and latency of the BOOM core.
The regression-based method \cite{lee2007regressDSE} (HPCA’07) leverages regression models with non-linear transformations to explore the power-performance Pareto curve.
We implement the key methods of these works and adopt them into SoC design exploration, then compare them with our framework. 
Moreover, we also compare SoC-Tuner with traditional multi-objective optimization methods like XGBoost\cite{chen2016xgboost}, random forest (RF), and support vector regression (SVR). 
Simulated annealing is leveraged for these traditional algorithms. 
We implement all methods in Python and all experiments are conducted on a Linux server with Intel Xeon CPU (E5-2630 v2@2.60GH) and 256 GB RAM.

\subsection{Experiment Setting}
We utilize the Chipyard to generate SoC hardware design as shown in \Cref{fig:soc-framework}. 
We use the ASAP7 process design kit (PDK) as the standard cell library and use tools integrated into Hammer \cite{liew2022hammer} to execute the VLSI flow.
The RTL-level simulation tool Verilator can obtain the accurate overall inference latency of various DNNs. 

For the experimental setting, we set $v_{th}=0.07$ for pruning design space, $u=0.1$, and $b=20$ for the SoC-Init algorithm. 
For a fair comparison, we keep the exploration iteration the same for baselines and our methods.  
All experiments and baselines are repeated $10$ times to get the corresponding average results.

In comparison with baselines, the average distance to reference set (ADRS) shown in Equation \eqref{equ:ards} is widely used in design space exploration to measure the distance between the learned Pareto optimal set with the real Pareto-optimal set of the design space. 
\begin{equation}
    ADRS(\Gamma,\Omega)=\dfrac{1}{|\Gamma|}\sum_{\gamma \in \Gamma,\omega \in \Omega} \min \quad f(\gamma,\omega),
    \label{equ:ards}
\end{equation}
where $f$ is the Euclidean distance function, $\Gamma$ is the real Pareto optimality set and $\Omega$ is the learned Pareto optimal set.

\subsection{Experimental Result and Analysis}


\Cref{fig:importance} demonstrates the result of importance analysis based on the ICD algorithm ($n=30$, $v_{th}=0.07$).
With the ICD algorithm, the whole design space points are pruned by about 30.16\%.

\minisection{Learned Pareto Optimal Set}.
We choose the benchmark ResNet50 as an example to show the superiority of SoC-Tuner in finding the Pareto optimal set of SoC design. In \Cref{fig:pareto}, gray points represents various design configuration of SoC, and colorful points are learned Pareto optimal set explored by SoC-Tuner and other methods. We only draw the design points close to the real Pareto optimal set to show the result clearly. 
Both in latency-area space shown in \Cref{fig:pareto-a} and latency-power space shown in \Cref{fig:pareto-b}, yellow circles represent the real Pareto design, and red diamonds represent the learned Pareto design by SoC-Tuner. 
The learned Pareto optimal set of SoC-Tuner is much closer to the real Pareto optimal set than other methods, demonstrating that SoC-Tuner's effectiveness outperforms other methods in finding the Pareto optimal set. 

Moreover, we use the simplified model\cite{samajdar2018scale-sim} to explore space according to its inaccurate metrics. 
To show the gap between the simplified model and RTL simulation, we simultaneously draw their learned Pareto optimal set in \Cref{fig:pareto-c}. The green hollow triangles represent Pareto optimal set found by the simplified model, while solid triangles represent the actual metrics from VLSI flow with the same design parameter.
\Cref{fig:pareto-c} proves that the simplified model cannot effectively guide the design of space exploration.

\begin{figure}[tb!]
    \centering
    \subfloat[]{
        \includegraphics[width=.235\textwidth]{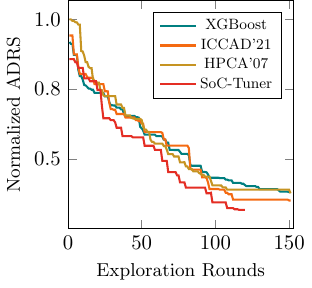}  \label{fig:results-a}
    } 
    \subfloat[]{
        \includegraphics[width=.235\textwidth]{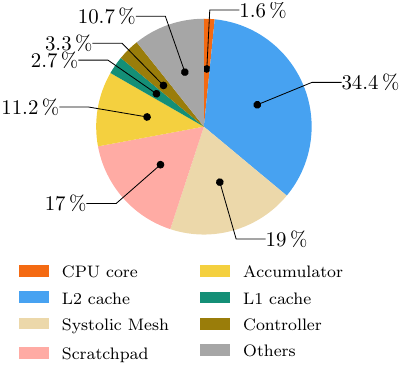}  \label{fig:results-c}
    }
    \caption{
        Experimental results.
        (a) ADRS curves of various methods;
        (b) Area breakdown of optimal SoC.
    }
    \label{fig:results} 
\end{figure}

\minisection{Learning Convergence and Optimal SoC Design}.
\Cref{fig:results-a} shows the ADRS in each exploration round. 
In each exploration round, SoC-Tuner outperforms previous methods. 
\Cref{fig:results-a} demonstrates that SoC-Tuner has higher exploration efficiency than other methods, giving a better SoC design in a shorter time.

\begin{figure}
    \centering
    \includegraphics[width=.46\textwidth]{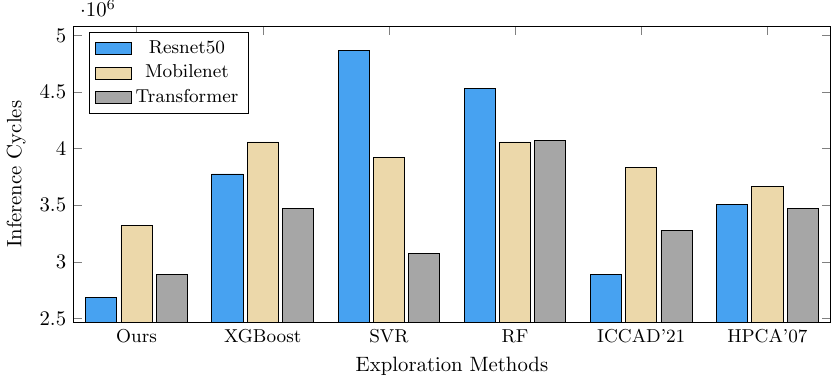}
    \caption{Inference cycles of DNNs on the learned optimal SoC designs obtained by different exploration methods.}
    \label{fig:results-b}
\end{figure}

 The optimal design points from the learned Pareto optimal sets given by various methods are listed in \Cref{fig:pareto}.
Since Transformer has too many parameters to be simulated in an acceptable time, we evaluate the inference latency on the $6$ basic structures, \textit{i.e.}, Decoder.
\Cref{fig:results-b} compares the inference cycles of optimal SoC designs explored by various methods, showing that the SoC designed by SoC-Tuner can get the least inference latency on various DNN workloads. 
The inference speed demonstrates our framework can find the optimal SoC design to obtain high performance in DNN acceleration. 
Moreover, our framework can facilitate SoC designers to design practical SoCs for DNN acceleration, instead of staying inaccurate simulation stage like previous simplified analytical tools.
With the learned Pareto optimal set, we can implement the optimal SoC design and \Cref{fig:results-c} shows the area breakdown of the design given by VLSI flow. 

%% file: doc/conc.tex
\section{Conclusion}\label{sec:conc}

In this paper, we have proposed SoC-Tuner, a novel exploration framework that utilizes a series of importance-guided algorithms to reduce the design iterations and find the Pareto optimal set of SoC configurations. 
Our framework thoroughly constructs a huge design space and analyzes the importance of design parameters in a typical DNN-targeting SoC.
The framework provides a group of efficient algorithms to prune the original design space and initialize the exploration.
Moreover, we utilize a novel multi-objective exploration with information gain to find the optimal SoC design for DNN accelerations. 
For designers, our framework can help them design A high-performance and low-cost SoC for various DNN applications on edge devices. 
For researchers, our framework brings more insights into the community of hardware design space. 
In future work, we plan to extend our framework to support more complicated DNN models like large language models and introduce more design constraints such as reliability, security, and robustness. 

\section*{Acknowledgements}

This research was partially supported by ACCESS -- AI Chip Center for Emerging Smart Systems, Hong Kong SAR.